\documentstyle[prl,twocolumn,epsf,floats,aps]{revtex}

\newcommand{\mean}[1]{\langle #1 \rangle}
\newcommand{\ave}[1]{\left\langle\!\left\langle #1 
\right\rangle\!\right\rangle} 
\newcommand{\til}[1]{{\tilde #1}}
\newcommand{\limit}{\lim_{{\scriptstyle n\to 0}\atop 
{\scriptstyle \bA\to 0}}}
\newcommand{\inter}{\leftrightarrow}
\newcommand{\bA}{{\bf A}}
\newcommand{\de}{\delta}

\newcommand{\adg}{a^{\dagger}}
\newcommand{\bdg}{b^{\dagger}}

\newcommand{\aph}{a^{\vphantom{\dagger}}}
\newcommand{\bph}{b^{\vphantom{\dagger}}}

\newcommand{\bd}{b^*}

\newcommand{\bh}{{\hat b}}
\newcommand{\bhd}{{\hat b}^*}
\newcommand{\bhdg}{{\hat b}^{\dagger}}

\newcommand{\be}{\begin{equation}}
\newcommand{\ee}{\end{equation}}
\newcommand{\bea}{\begin{eqnarray}}
\newcommand{\eea}{\end{eqnarray}}
\newcommand{\no}{\nonumber}
\newcommand{\br}{\no\\&&}

\newcommand{\tr}{{\rm tr}}
\newcommand{\Tr}{{\rm Tr}}
\newcommand{\ra}{\rangle}
\newcommand{\la}{\langle}

\newcommand{\intx}{\int_{0}^{L} \!\!\! dx}
\newcommand{\intt}{\int_{0}^{C} \!\!\! d\tau}
\newcommand{\intr}{\intx\!\intt}
\newcommand{\intu}{\int_{0}^{1} \!\!\! du}
\newcommand{\fintq}{\int\!\!D[Q]\,}

\newcommand{\dx}{\partial_x}
\newcommand{\dxprime}{\partial_{x'}}
\newcommand{\dt}{\partial_{\tau} \!}
\newcommand{\du}{\partial_u \!}
\newcommand{\dm}{\partial_{\mu} \!}
\newcommand{\Dm}{D_{\mu}}
\newcommand{\Dt}{D_{\tau}}

\begin{document}
\draft
\hyphenation{Fok-ker}
\hyphenation{Chal-ker}
\hyphenation{Doh-men}

\twocolumn[\hsize\textwidth\columnwidth\hsize\csname
@twocolumnfalse\endcsname

\title{Conductance and its universal fluctuations in the directed network 
model at the crossover to the quasi-one-dimensional regime} 
\author{Ilya A. Gruzberg and N. Read} 
\address{Department of Physics, Yale University  
P.O. Box 208120, New Haven, Connecticut 06520-8120\\
and Institute for Theoretical Physics, University of California 
Santa Barbara, California 93106-4030} 
\author{Subir Sachdev}
\address{Department of Physics, Yale University 
P.O. Box 208120, New Haven, Connecticut 06520-8120}
\date{June 10, 1997}
\maketitle

\begin{abstract}

The directed network model describing chiral edge states on the surface of
a cylindrical 3D quantum Hall system is known to map to a one-dimensional
quantum ferromagnetic spin chain. Using the spin wave expansion for this
chain, we determine the universal functions for the crossovers between 
the 2D chiral metallic and 1D metallic regimes in the mean and variance 
of the conductance along the cylinder, to first nontrivial order.

\end{abstract}

]

%%%%%%%%%%%%%%%%%%%%%%%%%%%%%%%%%%%%%%%%%%%%%%%
\section{Introduction}
\label{intro}

Mesoscopic disordered conductors have attracted many theoretical and
experimental investigations in recent years. The transport properties of
chiral edge states on the boundary of a two-dimensional quantum Hall state
have been of particular interest. Recently,
attention~\cite{cd,bafi,kim,bfz,m,y,grs} has focussed on a layered {\em
three}-dimensional (3D) system consisting a large number of such quantum
Hall systems stacked upon each other. Chalker and Dohmen~\cite{cd} argued
that the localized electronic states in the bulk of a quantum Hall system
will remain localized even in the presence of a coupling in the third
dimension, provided this coupling is not too strong. Electronic transport
in such a system is therefore controlled by the itinerant, chiral, edge
states, which now form a two-dimensional surface sheath on the boundary of
the 3D sample. A recent experiment~\cite{druist} on a 
GaAs/${\rm Al}_{0.1} {\rm Ga}_{0.9}$As multilayer structure has indeed
found that the conductivity perpendicular to the layers scales with the
perimeter of the sample when the bulk is in a layered quantum Hall state,
thus demonstrating the existence of a conducting surface sheath. 

\begin{figure}
\epsfxsize=3.4in
\centerline{\epsffile{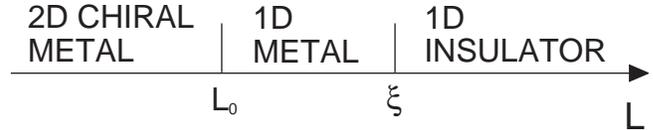}}
\vspace{0.2in}
\caption{Sketch of the crossovers as a function of the length $L$. The
first crossover~\protect\cite{bfz} is at the scale $L_0 = (C \rho_s / M_0
)^{1/2}$, where $\rho_s$ and $M_0$ are related to microscopic length and
energy scales in a manner discussed below (\protect\ref{contaction}) and
in Ref.~\protect\cite{grs}.  The second crossover is near the localization
length $\xi = 8 C \rho_s$.  The conditions for the validity of the
continuum theory require $\xi \gg L_0$, and so the two crossover scales
are well separated. The 2D chiral metal regime was called 0D in
Ref.~\protect\cite{bfz}, but we prefer the present terminology for reasons
discussed in Ref.~\protect\cite{grs}.}
\label{fig1}
\end{figure}

Theoretical analyses of transport in this chiral surface sheath have so
far been carried out in the context of a {\em directed network model} (DN)
introduced by Chalker and Dohmen~\cite{cd}. This model ignores
electron-electron interactions and describes the motion of independent
electrons along the links of a directed network; scattering events occur
at the nodes of the network and we are ultimately interested in the
probability distribution of transport coeffecients after the random
scattering matrix element has been averaged over. The global phase diagram
of the DN model is summarized in Fig~\ref{fig1}. 

We are considering a surface sheath of length $L$ perpendicular to the
layered quantum Hall states, and circumference $C$ along the chiral edges.
The system therefore has the geometry of a cylinder of length $L$ and
circumference $C$, and we discuss the results of two-terminal transport
measurements with leads placed at the ends of the cylinder. As a function
of $L$, there are three distinct coherent transport regimes, separated by
the lengths $L_0$ and $\xi$ at which there are smooth crossovers.  The
precise values of these lengths will be discussed later, but here we
simply note that $L_0 \sim (C a_x)^{1/2}$ and $\xi \sim C$, where $a_x$ is
a microscopic length of order the spacing between the layers. In the
presence of electron-electron interactions there will also be a
temperature-dependent phase coherence length $L_{\varphi}$ which arises
from inelastic scattering events between electrons. This length must be
accounted for in any comparison with experiments. In this paper we will,
for simplicity, not consider electron-electron interactions and, 
therefore, our results can only be applied when the temperature is 
sufficiently low so that $L_{\varphi} \gg L, C$. 

Let us now review the basic physical properties of the three regimes
in Fig~\ref{fig1}:
\newline
(i) {\em 2D Chiral Metal}: In the plane of the layers, the electrons move
ballistically in a single direction along the surface (the chiral edge
states), while transverse to the layers the motion is diffusive.  As the
transverse length $L < C^{1/2}$, the electrons do not have enough time to
execute one chiral orbit in the plane before exiting through one of the
ends of the system. The nomenclature ``2D chiral metal'' was suggested in
Ref~\cite{grs}, from an interpretation of the structure of conductance
fluctuations.  Earlier work~\cite{bfz} has referred to this regime as 0D. 
\newline
(ii) {\em 1D Metal}: The motion is as in the 2D chiral metal, 
but the system is now long enough that the electrons execute many chiral
transverse orbits before leaving through one of the ends of the cylinder.
The probability distribution of the transverse conductance is now very similar
to that of a short metallic wire~\cite{grs}.
\newline
(iii) {\em 1D Insulator}: If the cylinder is long enough, eventually the
one-dimensional diffusive motion along the axis of the cylinder undergoes
Anderson localization~\cite{cd}, and the system behaves like an insulating
wire~\cite{grs}. 

We now outline the results obtained in earlier studies of the DN model.
Ref~\cite{cd} presented numerical studies which obtained the crossover to
the 1D insulator.  A mapping of the DN model to a ferromagnetic spin chain
model was obtained in Ref~\cite{kim}. Ref~\cite{bfz} studied the energy
level fluctuations exclusively in the 2D chiral metal. In our previous
work~\cite{grs} (hereafter referred to as I), we obtained explicit results
for the crossovers in the conductance and its variance between the 1D
metal and the 1D insulator regimes: these results were in good agreement
with the earlier numerical work~\cite{cd}. Refs~\cite{m} and~\cite{y}
presented an incomplete description of the conductance fluctuations in the
2D chiral metal regime, and argued these were non-universal. 

The present paper contains a detailed and complete description of the
crossover between the 2D chiral metal and 1D metal, along with explicit
and exact expressions for universal crossover functions describing the
behavior of the conductance and its variance. In contrast to earlier
work~\cite{m,y} we find that the variance of the conductance, and all of
its higher moments, are universal when suitably expressed in terms of a
small number of renormalized parameters which characterize the microscopic
theory: the relationship between our and earlier~\cite{m,y} results will
be discussed in greater detail in the body of the paper. 

\section{Definition of the model and outline of results}
\label{outline}

The formalism and results of this paper are a continuation of those
presented in I. We will therefore not repeat here the detailed discussion
of the structure of the DN model presented in I. It was shown in I (and in
Refs~\cite{kim,bfz}) that the DN model can be mapped onto a continuum
quantum field theory for a 1D quantum ferromagnet in ``imaginary time''. 
Here the co-ordinate along the circumference of the cylinder (of length
$C$) is interpreted as the ``imaginary time'' direction of a
one-dimensional quantum system whose degrees of freedom reside along the
``space'' direction of length $L$. The mapping involves an average over 
disorder, which is performed using either replica or supersymmetry 
methods. In the first case the quantum ferromagnet contains spins of 
the su($n$,$n$) or su($2n$) algebra, depending on whether we use bosonic 
or fermionic replicas, and we are interested in the limit $n \to 0$. In the 
second case the spins are generators of a superalgebra. 
%Because the mapping involves an average over disorder, the quantum
%ferromagnet contains either $n$ replicated spins and we are interested in
%the limit $n \rightarrow 0$, or the spins are generators of a
%superalgebra. 
In the present paper, we will be interested in the perturbative spin-wave
expansion, and for this purpose it is more convenient to use a bosonic
replica formalism rather than supersymmetry. Although replicas are known
to fail in some non-perturbative situations~\cite{vz} (but not in all, see
for example Ref.~\cite{gm}), for perturbative calculations they are
completely equivalent to the supersymmetry method. If one uses, then, the
bosonic replica formalism to treat the disorder averages, the ferromagnet
action is (the reader is urged to consult our companion paper I for
further details)
\bea
S_{\rm cont} & = & -\intr\left( {M_0\over2} \intu \, \tr Q(u) \du Q(u)
\dt Q(u)\right. \no \\
& & \left.\vphantom{\intu}+ {\rho_s\over2} \tr (\dx Q)^2 \right),
\label{contaction}
\eea
where $M_0$ and $\rho_s$ are the magnetization per unit length, and the
spin stiffness in the ground state respectively. $Q(x,\tau)$ is a
$2n\times 2n$ matrix obeying $Q^\dagger=\Lambda Q \Lambda$, where
$\Lambda$ is a diagonal matrix with elements $(1,\ldots,1,-1,\ldots,-1)$,
$Q^2=I$, and having $n$ eigenvalues 1, $n$ eigenvalues $-1$, which thus
parametrizes the coset space SU($n$,$n$)/S(U($n$)$\times$U($n$)).  In the
Berry phase term, the first term in (\ref{contaction}), $Q(u) \equiv
Q(x,\tau,u)$ is some smooth homotopy between $Q(x,\tau,0) = \Lambda$ and
$Q(x,\tau,1) = Q(x,\tau)$.  The field $Q$ satisfies the periodic boundary
condition in the $\tau$-direction, and the following boundary conditions
at the spatial boundary, where the DN is connected to ideal leads: 
\be
Q(0,\tau,u) = Q(L,\tau,u) = \Lambda.
\label{bc}
\ee

This paper shall focus exclusively on the properties of $S_{\rm cont}$,
which the reader can also consider as a ferromagnet of super/replica spins
of interest in its own right. It was argued in I that the properties of
the DN model are completely, and universally, characterized by the
dimensionful parameters that appear in $S_{\rm cont}$: these are $\rho_s$,
$M_0$, $L$ and $C$. Of these, $L$ and $C$ are given by the macroscopic
dimensions of the sample, and therefore easily measured. The values of
$\rho_s$ and $M_0$ are determined by the detailed microscopic properties
of the electronic system; nevertheless if the microscopic Hamiltonian is
precisely known, the exact values of $\rho_s$ and $M_0$ can usually be
determined exactly~\cite{grs}. This is in contrast to the conventional
case in the theory of critical phenomena, where the relationship of
renormalized parameters to the underlying microscopics requires solution
of an intractable, strongly-coupled problem. 

We comment briefly on aspects of the relationship between the values of
$\rho_s$ and $M_0$ for the DN model as introduced in Ref.~\cite{cd}, and
studied in I. For the case in which the interior of the cylinder is
supposed to be a quantized Hall state with Hall conductance per layer
equal to 1 (in units of $e^2/h$), the parameter $M_0$ takes the
value~\cite{grs} $1/2a_x$, where $a_x$ is the separation of the layers,
while $\rho_s$ involves also other parameters that need not be specified
here~\cite{grs}. Here we wish to argue that, for the more general
situation in which the bulk Hall conductance is an integer $\nu>0$, say,
and there are $\nu$ chiral edge channels on the edge of each layer, the
appropriate model for the large scale properties is again the model with
action Eqn.~(\ref{contaction}), but with $M_0=\nu/2a_x$, where $a_x$ is
still the separation of the layers. This ensures that the edge channels
have the correct total conductance for transport along the edges, but the
details of the distinct channels are unimportant, assuming that there is
hopping of similar strength between all nearby channels. If the hopping
between the different edge channels in a layer is much weaker than that
between corresponding channels in different layers, then the theoretical
mappings of Refs.~\cite{kim,bfz} and I lead to $\nu$ ferromagnetic
superspin chains with weak ferromagnetic coupling between them. Such
couplings are relevant and cause the behavior to cross over on large
length scales to that for a single chain with $M_0=\nu/2a_x$. (The
relation among the values of $\rho_s$ for these different regimes is a
little more complex; it can be calculated by considering the excited
states of the coupled chains with a single spin flipped, in the long
wavelength limit.) It is possible that in practise a given sample with
$\nu>1$ might not be large enough to reach this regime, and the comparison
of experimental data with the following theory might then be complicated
by crossover effects. These crossover effects will not be further
addressed here, but could be studied by similar methods to those below. 

There are three distinct regimes in the theory described by $S_{\rm
cont}$, identified by Balents {\it et al.} in Ref.~\cite{bfz}, which 
were shown earlier in Fig~\ref{fig1}. We can now quote the precise
values of the scales $L_0$, $\xi$, at which the crossovers take place:
\be
L_0 = (C\rho_s/M_0)^{1/2}
\ee
and 
\be
\xi = 8C\rho_s. 
\ee
The physics of both 1D regimes was discussed in detail in I, where it was
shown that for $L\gg L_0$, the action for the quantum continuum
ferromagnet may be further reduced to that of a 1D non-linear sigma model. 
This latter model has been extensively studied in the context of quasi-1D
wires, see, in particular, Ref.~\cite{mmz}; using these earlier results,
explicit formulas for the mean and variance of the conductance of $S_{\rm
cont}$ were given in I, valid thoughout the crossover between the 1D
metallic and 1D localized regimes. 

In this paper we concentrate on the conduction properties in the 2D chiral
metal regime $L \ll L_0$ and the crossover to the 1D metallic regime
$L_0\ll L\ll \xi$. The model with action $S_{\rm cont}$ represents a
continuum su($n$,$n$) ferromagnet at finite effective temperature $1/C$
(this effective temperature, which is really the inverse circumference,
$1/C$, plays a very different role than the true temperature, which is
essentially zero, would). It was argued in Ref~\cite{bfz} and I, that when
the effective temperature $1/C$ becomes less than the order of the
low-lying level splittings of spin waves or magnons, which are of order
$\rho_s /M_0 L^2$, there are very few thermally excited magnons, and the
problem can be treated perturbatively. This defines the 2D chiral metal
regime, $1/C\ll\rho_s/M_0L^2$, or $L\ll L_0$. In the 1D metallic regime,
the temperature $1/C$ is larger than the splittings, but the thermally
excited magnons can be viewed as a slow variation of the spin direction
with position and imaginary time. In this regime, the classical
statistical mechanics approximation of neglecting the time dependence can
be used, and leads to the 1D non-linear sigma model, with coupling
constant $\sim 1/C\rho_s$. Perturbation theory breaks down on long length
scales, but is still valid when the length $L$ is less than the
localization length $\xi=8C\rho_s$. Thus, spin-wave perturbation theory
can be used all the way across the 2D to 1D crossover. In particular, the
scaling forms of the mean and variance of the conductance can be expanded
as a perturbation series in powers of $L/\xi$, times a universal function
of $L/L_0$ in each term. In this paper we will show explicitly that for
$\mean{g}$, the mean of $g$ (here and below, the single angular brackets
$\langle\ldots\rangle$ denote the average of a quantity over the
disorder), this takes the form (here and henceforth, all conductances are
quoted in units of $e^2 / h$): 
\be
\mean{g} = \xi/2L + (L/\xi) \Phi \left(L/L_0\right) + 
O\left((L/\xi)^2\right).
\label{meang} 
\ee
To leading order in $L/\xi$ there is no dependence on $L/L_0$
(Ref.~\cite{cd}).  We will show below that the possible term of order
$(L/\xi)^0$ vanishes identically, consistent with the known result that
the leading ``weak localization'' correction to $\mean{g}$ vanishes in the
quasi-1D metallic regime $L/L_0 \to \infty$ in the present (unitary) case. 
The next term in the expansion in $L/\xi$ in~(\ref{meang}) does have a
non-trivial crossover at the scale $L_0$, described by a universal
function $\Phi$ given below in Eq.~(\ref{Phi}). In contrast, for the
variance of $g$ we obtain
\be {\rm var} \, g \equiv \mean{g^2}-{\mean{g}}^2= \widetilde{\Phi} 
\left(L/L_0\right) + O\left(L/\xi\right),
\label{varg}
\ee
where the universal crossover from the 2D to the 1D metallic regime is
evident in the leading term $\widetilde{\Phi}$ given below in
Eq.~(\ref{Phitilde}). This leading term, which is all that survives in the
limit $L/\xi\rightarrow 0$, corresponds to what are known as ``universal
conductance fluctuations'' (UCF), which are similarly the weak coupling
limit of the crossover function. Note that we use the term ``universal''
as it is understood in critical phenomena, to mean that the results are
independent of microscopic details of the model, and that the conductance
fluctuations are described by a universal function of the sample geometry
(that is, in the present case, of $L/L_0$). This should be contrasted with
the usage often implicit in the mesoscopic physics literature, in which
``universal'' stands for the {\em value} $e^2/h$. In such usage, anything
close to this value is called universal.  For us, on the other hand, this
unit of conductance is left implicit in our formulas, and it is the
numerical coefficient, which is actually a precise function of $L/L_0$,
that is universal in our sense, and need not even be close to unity, as in
the results below. These considerations apply unchanged to the
conventional UCF, which in our opinion are best regarded in this sense
also. 

\begin{figure}
\epsfysize=4truein
\epsfxsize=4truein
\vspace{-.41cm}
\centerline{\epsffile{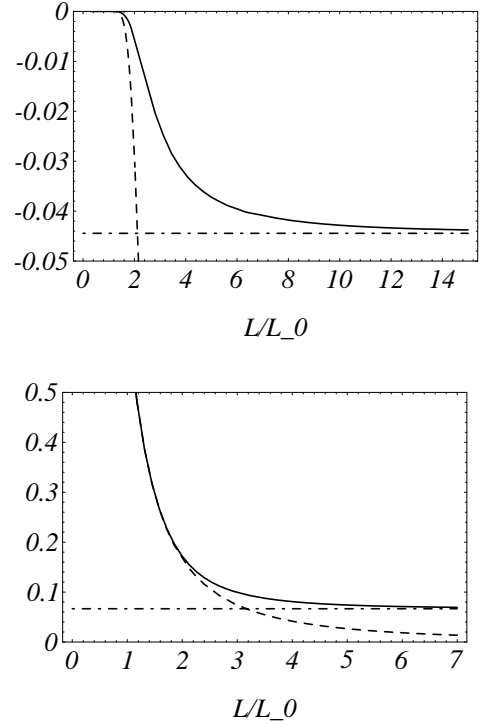}}
\caption{Plots of the universal functions $\Phi$, for the correction to
the mean conductance (top), and $\widetilde{\Phi}$, for the variance of
the conductance (bottom). The dashed and dot-dashed lines are plots of the
asymptotic expressions at small and large values of $L/L_0$,
Eqs.~(\protect\ref{Phi2},\protect\ref{Phitilde2})
and~(\protect\ref{Phi1},\protect\ref{Phitilde1}), respectively.}
\label{fig2} 
\end{figure}

The universal functions $\Phi$ and $\widetilde{\Phi}$ are calculated in
the following sections, and the results are
\bea
\Phi\left(L/L_0\right) & = & {4\lambda^2\over \pi^4} 
\left(\left(\sum_{l=1}^{\infty} f(\lambda l^2) \right)^2 + 
{1\over 2} \sum_{l=1}^{\infty} f^2(\lambda l^2) \right.\no \\
&&\mbox{} + \!\! \sum_{l_1,l_2,l_3 = 1}^{\infty} {l_3^2\over l_1^2 + l_2^2 - 
l_3^2}{f(\lambda l_1^2) f(\lambda l_2^2) f(-\lambda
l_3^2) \over f\left(\lambda (l_1^2 + l_2^2 - l_3^2)\right)} \no \\
&&\left.\vphantom{\left(\sum_{l=1}^{\infty}\right)^2}
\times \left(\de_{l_1,l_2+l_3} + \de_{l_2,l_1+l_3} + 
\de_{l_3,l_1+l_2} \right) \right),
\label{Phi} \\
\widetilde{\Phi}\left(L/L_0\right) & = & {\lambda\over 3\pi^2} +
{2\lambda\over \pi^4} \sum_{l=1}^{\infty} f(\lambda l^2)\!
\left(2/l^2 - \lambda f(-\lambda l^2) \right).
\label{Phitilde}
\eea
In these expressions $\lambda = 2\pi^2 L_0^2/L^2$, and $f(x) = 1/(e^{x}
-1)$ is the Bose function. These expressions are easily evaluated
numerically, and the resulting plots are shown in Fig.~\ref{fig2}. 

Using the limiting behaviour of $f(x)$ for small and large values
of $x$ we obtain from Eqs.~(\ref{Phi},\ref{Phitilde}) in the 1D limit 
$L/L_0\rightarrow\infty$ (but still with $L/\xi\ll 1$) of long cylinders 
\bea
\Phi \left(L/L_0\right) & \to & -{2\over 45}, \label{Phi1} \\
\widetilde{\Phi} \left(L/L_0\right) & \to & {1\over 15}.
\label{Phitilde1}
\eea
These limits~(\ref{Phi1},\ref{Phitilde1}) for long cylinders are the well
known quasi-1D metallic results~\cite{mmz}.
In the opposite 2D chiral limit $L/L_0\rightarrow 0$ of short cylinders
\bea
\Phi \left(L/L_0\right) & \sim & -{16 L_0^4\over L^4}
\exp\left(-{2\pi^2 L_0^2 \over L^2}\right), \label{Phi2} \\
\widetilde{\Phi} \left(L/L_0\right) & \sim & 
{2L_0^2\over 3L^2}. 
\label{Phitilde2}
\eea

As an alternative to the interpretation above in terms of thermally
excited magnons, we can also understand the results for this crossover in
terms of the sum over paths taken by the propagating electrons. To
calculate the conductance, retarded and advanced paths are required, and
these are paired up by the average over the disorder. This was the basis
for the treatment in I. For $\mean{g}$, the leading term comes from paired
paths that go from one end to the other of the system while propagating a
distance of order $M_0L^2/\rho_s$ around the circumference. This distance
is $\ll C$ in the 2D regime. Therefore the corrections (which correspond
to thermally-excited magnons) in the series in $L/\xi$ for $\mean{g}$,
which reflect the effect of paired paths that wind around the finite
circumference and interfere with themselves, are exponentially small in
this limit. In contrast, for ${\rm var}\, g$, there are nontrivial effects
at leading (zeroth) order in $L/\xi$, which, like the leading term in
$\mean{g}$, do not require winding paths. The correction terms to the
behavior in Eq.~(\ref{Phitilde2}) are also exponentially small in the 2D
chiral metal regime. Note that nontrivial effects in the absence of
winding paths also appeared in Refs.~\cite{mmz,m}. 

A result of the form of Eq.~(\ref{Phitilde2}) for ${\rm var} \, g$, has
been claimed by Mathur~\cite{m} and Yu~\cite{y}, who find ${\rm var} \, g
\propto C \rho_s/M_0 L^2$ in the limit $L \ll L_0$. Mathur's explanation
of this result, which is essentially that the sample can be divided into
blocks of size of order $L\times M_0L^2/\rho_s$ which have
statistically-independent fluctuations with variance of order $1$ that add
to give ${\rm var}\, g$, appears correct. However, both he and Yu~\cite{y}
state that this result is nonuniversal, and Yu bases this statement on its
dependence on $\rho_s/M_0$ (in our notation). Mathur's results are
incomplete, as stated by him. His result for $L\ll L_0$ (but not that for
$L\gg L_0$) does agree with ours in this limit for fortuitous reasons
which will be discussed in Sec. V.  Yu~\cite{y} states that his results
are nonperturbative, but fail for some parameter values. In contrast, we
are able not only to compute the mean and variance of the conductance in
the limiting cases, but also to determine the full crossover functions. We
have already commented above on the meaning of universality and the
scaling with $L/L_0$, and we argue that the crossover functions are
universal in the domain of applicability of the continuum description of
the DN model. Our calculation shows that this crossover from 2D to 1D
metal can be treated perturbatively in close analogy to the usual UCF in
an isotropic diffusive system, where one can also treat the crossover from
2D to 1D. (Here we would like to repeat a remark already made in I, that
in the latter case, for a sample of width $W$ and length $L$, one finds
${\rm var}\, g \propto W/L$ as $W/L\rightarrow \infty$. This behavior can
be understood by viewing the sample as composed of independent blocks of
size $L\times L$. It is nonetheless universal in the sense we prefer, as
discussed above. It is true that here there is no dependence on the
conductivity, while there is in the chiral metal, but this does not mean
the result for the latter is nonuniversal, because the scaling functions
for the chiral metal exhibit no-scale-factor universality, as argued in I,
so the dependence on the bare coupling constants $\rho_s$ and $M_0$ is
meaningful.) In addition, our results differ quantitatively from those of
Yu~\cite{y}. Thus our results differ from those in either of
Refs.~\cite{m,y}. 

The rest of the paper presents the details in the computation of the
central results (\ref{Phi}) and (\ref{Phitilde}).  Beginning in the next
section, we develop the spin wave expansion mentioned above. The
su($n$,$n$) spin chain is described in Sec.~\ref{DM}. In the same section
we introduce a parametrization for the spins in terms of Dyson-Maleev (DM)
bosons. In Sec.~\ref{gaugeinv} we exploit gauge invariance to obtain
expressions for the currents and for the moments of the conductance in
terms of DM bosons. Details of the perturbative calculations are given in
Sec.~\ref{perttheory}. Finally, we conclude in Sec.~\ref{conclusion}. 

%%%%%%%%%%%%%%%%%%%%%%%%%%%%%%%%%%%%%%%%%%%%%%
\section{Spin wave expansion in terms of Dyson-Maleev bosons}
\label{DM}

In this section we will begin by considering a lattice discretization of
$S_{\rm cont}$ and then set up a spin-wave perturbation theory using the
Dyson-Maleev method. 

In I we used supersymmetry to treat the disorder averages in the mapping
of the DN to a spin chain. When the mapping is performed instead (in a
completely parallel way) using bosonic replicas we obtain the following
Hamiltonian for the spin chain: 
\be
{\cal H} = K \sum_{i=1}^{N-1} \tr \left( J(i) J(i\!+\!1) + 
{1 \over 2}\Lambda \left(J(1) + J(N)\right)\right).  
\label{hamiltonian} 
\ee 
Here the $J$ are the su($n$,$n$) spins in a particular irreducible
representation.  This representation naturally appears in the mapping as
follows. We introduce $n$ ``retarded'' and $n$ ``advanced'' bosons $a_i,
b_i$, $i = 1,2,\ldots,n$. Their bilinear products obeying u($n$,$n$)
commutation relations are arranged in a $2n\times 2n$ matrix
\bea 
\label{Snn} 
J & = & \left( \begin{array}{cc} 
{1\over2}\delta_{ik} + \adg_i a_k & \adg_i \bdg_l \\ 
-\bph_j \aph_k & {1\over2}\delta_{jl} - \bph_j \bdg_l 
\end{array} \right)  
\eea
The average over randomness in the DN model produces the local constraint
$\adg_i \aph_i = \bdg_i \bph_i$ (we assume summation over repeated indices
everywhere, unless stated otherwise). The subspace of the Fock space of
bosons specified by this constraint forms a highest weight irreducible
representation of the algebra su($n$,$n$) with the vacuum $|0\ra$ being
the highest weight state. 

The trace in Eq.~(\ref{hamiltonian}) is over the matrix indices of the
spins which are multiplied as matrices. The last term in
Eq.~(\ref{hamiltonian}) comes from the boundary, where the DN is connected
to the ideal leads. $\Lambda$ is the same as in~(\ref{bc}). 

The mean conductance $\mean{g}$ is given by the thermal correlator of
conserved currents 
\be
\mean{g} = -\Tr \left( {\cal T} I(i) I(j) e^{-C{\cal
H}}\right), 
\ee
for $i\neq j$, where $\rm Tr$ stands for the quantum mechanical trace in
the Hilbert space, the role of the inverse temperature is played by the
circumference $C$ of the cylinder, the currents $I(i)$ are related to the
spins (see I), and $\cal T$ is the time-ordering operator. Similar
expressions can be given for the moments of the conductance. 

For the perturbative spin wave expansion it is convenient to make use of
another construction of the necessary representation of su($n$,$n$). It is
very similar to the Dyson-Maleev construction~\cite{dm} of representations
of su($2$) and proceeds as follows. We introduce $n^2$ bosons, which we
arrange in an $n\times n$ matrix $\bh$ with elements $\bh_{ij}$, $i,j =1,
\ldots, n$. To avoid confusion, we will denote boson creation operators by
an asterisk: $\bhd_{ij}$, and reserve the dagger for the hermitean
conjugate of matrices of operators. Then the hermitean conjugate matrix
$\bhdg$ has elements $(\bhdg)_{ij} = \bhd_{ji}$. We also assume that
whenever we write a product of several $\bh$'s and $\bhdg$'s without
indices, it means the usual matrix product. These $n^2$ bosons satisfy
canonical commutation relations $[\bh^{\phantom{*}}_{ij}, \bhd_{kl}] =
\delta_{ik} \delta_{jl}$. Then the su($n$,$n$) spin $J$ is given in
block-matrix form by
\be
\label{DMnn1}
J = \left( \begin{array}{cc} {1\over2} + \bhdg \bh & \bhdg \\
(n - 1 - \bh\bhdg)b & n -{1\over2} - \bh\bhdg \end{array} \right).
\ee

When we substitute this expression for the spins $J$ in terms of DM bosons
into the Hamiltonian~(\ref{hamiltonian}) and bring them to normal order,
we obtain, in particular, quartic terms which do not have the matrix product
index structure. Next we introduce a bosonic functional integral using 
bosonic coherent states $|b \ra$ satisfying $\bh|b \ra = b |b \ra$. In 
this functional integral the bosonic operators $\bh$ are replaced by 
commuting variables $b$ (without a hat), and we can bring the 
quartic terms back to the usual matrix product form. This gives the 
following action: 
\bea
S &=& \intt \sum_i \tr \left( \bdg(i)\dt b(i) \right.\no \\ 
&& \mbox{}+ K\left(\bdg(i\!+\!1)-\bdg(i)\right) \left(b(i\!+\!1)-b(i)\right) 
\no \\ && \left.- K\bdg(i) \left(b(i\!+\!1)-b(i)\right) \bdg(i\!+\!1)
\left(b(i\!+\!1)-b(i)\right)\!\right) \no 
\eea
(we dropped a constant $n/2$ which disappears in the replica limit).
The measure for the functional integral over $b$, $\bdg$ is the usual one, 
$D[b,\bdg]$. 

Next we take continuum limit in the spatial direction and obtain the
action $S=S_0+S_{\rm int}$, where
\bea
S_0 &=& \intr\,\tr\left(2M_0\bdg\dt b+4\rho_s\dx\bdg\dx b\right), 
\label{freeaction} \\
S_{\rm int} &=& -4\rho_s \intr\,\tr\left(\bdg\dx b\bdg\dx
b\right),
\label{interaction}
\eea
with $M_0=1/2a_x$ and $\rho_s=Ka_x$. The $\tau$-derivative term in $S_0$
does not have the canonical form. This could be mended by a rescaling of
the bosonic fields. However, we will keep the present normalization to
avoid cumbersome coefficients in different expressions appearing later.
The boundary terms in the Hamiltonian~(\ref{hamiltonian}) in the continuum
description force the field $b(x)$ to take zero values at the boundaries:
\be
b(0,\tau)=b(L,\tau)=0.
\label{bc1}
\ee

Before giving the expressions for the currents in terms of DM bosons and
evaluating their correlators in the theory with the
action~(\ref{freeaction},\ref{interaction}), we note that the latter could
have been obtained from $S_{\rm cont}$ of Eq.~(\ref{contaction}) by the
following formal change of variables in the functional integral over the 
field $Q$:
\be
Q = \left( \begin{array}{cc} 1 + 2 b \bdg & 2\bdg \\
-2(1 + b \bdg)b & - 1 - 2 b \bdg \end{array} \right).
\label{changevar}
\ee
Even though this expression does not satisfy the original conditions on 
$Q$ (stated after Eq.~1), it formally coincides with $2\la 
b|J|b\ra$ where $J$ is given by~(\ref{DMnn1}) ($n$ in the lower left 
corner of~(\ref{DMnn1}) disappears after normal ordering the term 
$\bh\bhdg$). Also, the SU($n$,$n$)-invariant functional measure for $Q$ 
reduces to $D[b,\bdg]$. These observations give us a very general and 
straightforward way of obtaining the expressions for the currents and 
the moments of the conductance. Namely, we have to make the 
action~(\ref{contaction}) gauge invariant by coupling to a gauge field. 
Then functional derivatives with respect to this field will give the 
currents and the moments of the conductance. When used together with 
Eq.~(\ref{changevar}), this procedure gives all the quantities in terms 
of DM bosons. In the next section we describe this procedure in detail.

%%%%%%%%%%%%%%%%%%%%%%%%%%%%%%%%%%%%%%%%%%%%%
\section{Gauge invariance, current conservation and moments of
conductance}
\label{gaugeinv}

It is clear from the last section that the distribution of the conductance
of the DN model is related to correlators 
of the conserved
current of the ferromagnetic spin chain.
The computation of the required correlators
is greatly simplified by an understanding of the
gauge-invariance properties of $S_{\rm cont}$ which 
will be discussed in this section.

We can make the action (\ref{contaction}) invariant under the local
gauge transformation $Q \to \til{Q} = U Q U^{-1}$, $A_{\mu}\to\til{A_{\mu}} 
= U A_{\mu} U^{-1} - i U \dm U^{-1}$, with $U \in$ SU($n$,$n$), by 
replacing the partial derivatives with the covariant ones: $\dm X \to 
\Dm X \equiv \dm X + i [A_{\mu},X]$, where the gauge potential $A_{\mu}$ 
is an element of su($n$,$n$). However, as discussed in detail in 
Ref.~\cite{xrs}, this is only necessary for the $(\dx Q)^2$ term. The Berry phase 
term, being a total derivative, can be made gauge invariant by adding a 
boundary term, and the gauge invariant action is 
\bea
S[\bA] & = &  -\intr\left( {M_0\over2} \intu \, \tr Q(u) \du Q(u) \dt
Q(u) \right.\no \\
& &\left.\vphantom{\intu} + i\, M_0 \tr A_{\tau} Q + {\rho_s\over2} \tr 
(D_x Q)^2 \right). \label{giaction}
\eea

Gauge invariance of this action implies some conservation laws, or Ward
identities (WI's). In their derivation we closely follow Ref.~\cite{xrs}. 
Let us introduce the generating functional 
\be
Z[\bA] = \fintq
\exp\left(-S[\bA]\right).
\ee  
Using the notation 
\be
\ave{X} =
Z^{-1}[\bA]\fintq X[Q] \exp\left(-S[\bA]\right),
\ee
where $X[Q]$ represents any functional of $Q$, we can show that gauge 
invariance of the action
(\ref{giaction}) results in the following equation of motion: 
\be
\ave{\de S
/ \de Q} = 0.
\ee 
As shown in Ref.~\cite{xrs}, this is equivalent to the following
covariant current conservation law: 
\be
\Dt\ave{j_{\tau}} + D_x\ave{j_x} = 0,
\label{wi}
\ee
where we introduced matrix-valued gauge invariant currents $j_{\mu}$ with
elements
\be
j_{\mu, ij}(x,\tau) \equiv - {\de S[\bA] \over \de A_{\mu,
ji}(x,\tau)}.
\label{defcurrents}
\ee

Equation~(\ref{wi}) is the first in the series of WI's coming
from the gauge invariance. The gauge invariant currents
$j_{\mu}$~(\ref{defcurrents}) contain gauge potentials, and further
WI's are obtained differentiating with respect to $\bA$.
In particular, we can obtain such identities for the mean two-pro\-be
conductance and its variance, which in this field-theoretic formalism are
obtained as follows. Let us assume that the source field $\bA$ is
independent of $\tau$, and moreover, has the special form
\be
A_{\tau} = 0, \quad A_x(x) = \left( \begin{array}{cc} A^{(1)}(x) & A^+(x) \\
A^-(x) & A^{(2)}(x) \end{array} \right).
\label{specpoten}
\ee
Then the mean conductance is given by
\be
\mean{g(x,x')} = \limit {\de^2 Z[\bA]\over
\de A^+_{11}(x) \de A^-_{11}(x')}.
\label{meang1}
\ee
If we introduce the spatial currents integrated over $\tau$,
\be
I_{ij}(x) = - {\de S[\bA]\over \de A_{ji}(x)} =
\intt j_{x,ij}(x,\tau),
\label{current}
\ee
then
\be
\mean{g(x,x')} = \limit {\de\ave{I^+_{11}(x')}\over\de A^+_{11}(x)}.
\label{meang2}
\ee
This differs from the formula for $\mean{g}$ earlier by including a 
contact term, and which was avoided then by the condition on the sites
$i \neq j$.

With the choice~(\ref{specpoten}) the WI~(\ref{wi}) becomes
\be
D_x\ave{I(x)} = \dx\ave{I(x)} + i[A_x,\ave{I(x)}] = 0.
\label{wi1}
\ee
The current $I(x)$ has the same structure as $A(x)$:  
\be
I(x) = \left( \begin{array}{cc} I^{(1)}(x) & I^+(x) \\
I^-(x) & I^{(2)}(x) \end{array} \right),
\ee
and the $+$ component of~(\ref{wi1}) is simply 
$$
\dx\ave{I^+} \!+\! i \left\la \!\!\left\la A^+ I^{(2)} \!-\! I^{(1)} A^+ 
\!+\! A^{(1)}I^+ \!-\! I^+ A^{(2)} \right\ra \!\!\right\ra  = 0. 
$$
Differentiating with respect to $A^+_{11}(x')$ and using~(\ref{meang2}), 
we obtain 
$$
\dx\mean{g(x',x)} + i \de(x-x') \limit \left\la\!\!\left\la 
I^{(2)}_{11}(x) - I^{(1)}_{11}(x)\right\ra\!\!\right\ra = 0. 
$$
In the next section we will show that the second term here vanishes in 
all orders of perturbation theory. As a consequence, the mean conductance is 
independent of the positions of cross section $x$. Similarly, it is 
independent of $x'$ as well: 
\be
\dx\mean{g(x,x')} = \dxprime\mean{g(x,x')} = 0.
\label{wicond}
\ee
This can be used to average $\mean{g}$ over these positions: 
\bea
\mean{g} & = & {1\over L^2} \intx\intx' \mean{g(x,x')} \label{meang3} \\
& = & {1\over L^2} \limit\intx\intx' 
\left\la\!\!\!\left\la{\de I^+_{11}(x') \over \de A^+_{11}(x)} +
I^-_{11}(x) I^+_{11}(x')\right\ra\!\!\!\right\ra. \no
\eea

Similarly, the second moment is obtained by taking four derivatives:
\bea
& & \mean{g(x_1,x'_1)g(x_2,x'_2)} = \no \\
& & \limit {\de^4 Z[\bA] \over \de A^+_{11}(x_1) \de A^-_{11}(x'_1)
\de A^+_{22}(x_2) \de A^-_{22}(x'_2)}.
\eea
Again we can show that this expression does not depend on any $x_i$, and
we write
\bea
{\rm var\,}g & = & {1\over L^4}
\limit\intx_1\!\!
\intx'_1\!\! \intx_2\!\! \intx'_2 \no \\
& & \left\la\!\!\!\left\la
{\de I^+_{11}(x'_1) \over \de A^+_{11}(x_1)}
{\de I^+_{22}(x'_2) \over \de A^+_{22}(x_2)} +
{\de I^+_{22}(x'_2) \over \de A^+_{11}(x_1)}
{\de I^-_{22}(x_2) \over \de A^-_{11}(x'_1)}\right.\right. \no \\ & & +
{\de I^-_{22}(x_2) \over \de A^+_{11}(x_1)}
{\de I^+_{22}(x'_2) \over \de A^-_{11}(x'_1)} \no \\ & & +
{\de I^+_{11}(x'_1) \over \de A^+_{11}(x_1)}
I^-_{22}(x_2) I^+_{22}(x'_2)  + (1\inter 2) \no \\  & & +
{\de I^+_{22}(x'_2) \over \de A^+_{11}(x_1)}
I^-_{22}(x_2) I^+_{11}(x'_1)  + (1\inter 2) \no \\  & & +
{\de I^-_{22}(x_2) \over \de A^+_{11}(x_1)}
I^+_{11}(x'_1) I^+_{22}(x'_2) + (+\inter -) \no \\  & & 
+\left.\left.\vphantom{\de I^+_{11}(x'_1) \over \de A^+_{11}(x_1)}
I^-_{11}(x_1) I^+_{11}(x'_1) I^-_{22}(x_2) I^+_{22}(x'_2)
\right\ra\!\!\!\right\ra_{\rm conn},
\label{varg1}
\eea
where the subtraction of $\mean{g}^2$ leaves only connected terms in the
above expression.

Now we specify the gauge potential even further, leaving only the
$A^{\pm}$ components nonzero, which is all we need to calculate moments of
conductance. With this gauge choice we make the
substitution~(\ref{changevar}) in the gauge invariant action $S[\bA]$. The
result is $S[\bA] = S_0 + S_{\rm int} + S'[A_x]$, where $S_0$ and $S_{\rm
int}$ are the same as before, Eqs.~(\ref{freeaction},\ref{interaction}),
and
\bea
S'[A_x]&=&-4\rho_s\intr\,\tr\left(iA^-(\dx\bdg-2\bdg\dx 
b\bdg)\right.\no\\
&& \mbox{}+ iA^+\left(\dx(b+b\bdg b)-2b\dx\bdg b+2 b\bdg\dx b\bdg b\right)\no\\
%&& + iA^{(1)}(\dx \bdg b - \bdg \dx b - 2 \bdg \dx b \bdg b) \no \\
%&& + iA^{(2)}(\dx b\bdg - b\dx\bdg + 2 b\bdg\dx b\bdg) \no \\
&& \mbox{}+ A^+ A^- +A^+ A^-\bdg b +A^+ b\bdg A^- \no\\
&& \mbox{}+ 2A^+ b\bdg A^-\bdg b - A^-\bdg A^-\bdg -A^+ b A^+ b\no\\
&& \left.- 2A^+ b A^+ b\bdg b -A^+ b\bdg b A^+ b\bdg b \right) .
\label{sprime}
\eea

The currents $I_x^{\pm}$ and their derivatives necessary for the
calculation of $\mean{g}$ and ${\rm var}\, g$ follow from~(\ref{current}) 
and~(\ref{sprime}) (we assume notation ${\bf r} \equiv (x,\tau)$,
${\bf r'} \equiv (x',\tau')$, etc.):
\bea
\left.I^+_{ij}(x)\right|_{\bA=0} & = &-4i\rho_s \intt 
\left(\dx\bdg - 2\bdg \dx b\bdg \right)_{ij}({\bf r}), \no \\
\left.I^-_{ij}(x)\right|_{\bA=0} & = & -4i\rho_s \intt \left(\dx (b+b\bdg 
b) - 2 b\dx\bdg b \right.\no \\
&& \left.+ 2b\bdg\dx b\bdg b \right)_{ij}({\bf r}), \no \\
{\de I^+_{ij}(x') \over \de A^+_{kl}(x)} & = & -4\rho_s \de(x-x') \intt
\left(\de_{ik} \de_{jl} + (\bdg b)_{ik}\de_{jl} \right.\no \\
&& \left.+ \de_{ik}(b\bdg)_{lj} + 2 (\bdg b)_{ik}(b\bdg)_{lj} \right)({\bf 
r}), \no \\
{\de I^-_{ij}(x') \over \de A^-_{kl}(x)} & = & -4\rho_s \de(x-x') \intt
\left(\de_{ik} \de_{jl} + (b\bdg)_{ik}\de_{jl}\right. \no \\
&& \left.+ \de_{ik}(\bdg b)_{lj} + 2 (b\bdg)_{ik}(\bdg b)_{lj} \right)({\bf 
r}), \no \\
{\de I^+_{ij}(x') \over \de A^-_{kl}(x)} & = & 8\rho_s \de(x-x') \intt
\left(\bd_{ki}\bd_{jl}\right)({\bf r}), \no \\
{\de I^-_{ij}(x') \over \de A^+_{kl}(x)} & = & 8\rho_s \de(x-x') \intt
\left(b_{ik}b_{lj} + b_{ik}(b\bdg b)_{lj} \right.\no \\
&& \left.+ (b\bdg b)_{ik}b_{lj} + (b\bdg b)_{ik}(b\bdg b)_{lj}\right)({\bf 
r}). \label{currents}
\eea

If we use all four blocks of $A_x$ in this derivation, we would also 
obtain the following expressions for the diagonal currents, necessary to 
prove Eq.~(\ref{wicond}):
\bea
\left.I^{(1)}_{ij}(x)\right|_{\bA=0} & = &-4i\rho_s \intt \left(
\dx \bdg b - \bdg \dx b\right. \no \\
&& \left.- 2 \bdg \dx b \bdg b \right)_{ij}({\bf r}), \no \\
\left.I^{(2)}_{ij}(x)\right|_{\bA=0} & = & -4i\rho_s \intt \left(
\dx b\bdg - b\dx\bdg \right.\no \\
&&\left. + 2 b\bdg\dx b\bdg \right)_{ij}({\bf r}).
\label{diagcurrents}
\eea

Substituting expressions~(\ref{currents}) into 
Eqs.~(\ref{meang3},\ref{varg1}),
we notice that the total derivative terms in $I^{\pm}$ give zero after
integration over $x$'s, thanks to the boundary conditions (\ref{bc1}).
Dropping these terms, we obtain the exact expression for the mean 
conductance \bea
\mean{g} & = & {4C\rho_s\over L} + {4\rho_s\over L^2} \intr \no \\
&&\left\la\!\left\la(\bdg b)_{11} + (b\bdg)_{11} + 2 (\bdg 
b)_{11}(b\bdg)_{11} \right\ra\!\right\ra \no \\
&&\mbox{} - {64\rho_s^2\over L^2} \left\la\!\!\!\left\la \intr (b\dx\bdg b - 
b\bdg\dx b\bdg b)_{11}({\bf r}) \right.\right.\no \\
&& \left.\left.\times \intx'\!\!\intt' (\bdg \dxprime b 
\bdg)_{11}({\bf r'}) \right\ra\!\!\!\right\ra \label{meang4} \eea
(from here on, the limits $n\rightarrow 0$ and ${\bf A}\rightarrow 0$ 
will be implicit in expressions for observables such as $\mean g$ and 
${\rm var}\,g$).
The expression for ${\rm var}\, g$ is more complicated and we do not 
reproduce
it here. It will be clear from the next section, where we discuss the
details of perturbative calculation, that only two terms in the above
expression (\ref{meang4}) contribute to the function $\Phi$ of
Eq.~(\ref{meang}).  Similar simplifications are even more drastic for the
variance of the conductance. 

%%%%%%%%%%%%%%%%%%%%%%%%%%%%%%%%%%%%%%%%%%%%%%
\section{Details of perturbative calculation}
\label{perttheory}

We have now assembled all the tools required to compute the mean conductance
and its variance, and are ready to perform the explicit computation
of the universal functions $\Phi$ and $\tilde{\Phi}$.

The free action $S_0$ is diagonalized by Fourier transformation of the
fields $b({\bf r})$ to momentum space $b({\bf r}) = C^{-1} \sum_{k,m}
b({\bf p}) \phi_k(x) e^{-i\omega_m \tau}$, where ${\bf p} \equiv 
(k, i\omega_m)$ and $\phi_k(x) = (2/L)^{1/2} \sin kx$:
\be
S_0 = {1\over C} \sum_{k,m}\left(4\rho_s k^2 - 2i M_0 \omega_m\right) 
\tr\bdg({\bf p}) b({\bf p}).
\label{freeaction1}
\ee
Periodic boundary conditions in the $\tau$-direction and the homogeneous
ones in the $x$-direction (\ref{bc1}) imply that the frequencies
$\omega_m$ and momenta $k$ take the values
\bea
\omega_m & = & {2\pi m/C}, \quad m = 0,\pm1,\ldots, \no \\
k_l & = & {\pi l/L}, \quad l = 1,2,\ldots
\eea

The action~(\ref{freeaction1}) implies that the bare propagator of the bosons
$b$ (analogous to the diffuson of the usual isotropic metallic systems) is 
diagonal in replica indices: 
\be
\left\la\!\left\la\bd_{pq}({\bf p}) b_{rs}({\bf p'}) \right\ra\!\right\ra_0 =
\de_{pr} \de_{qs} \de_{{\bf p}{\bf p'}} d_0({\bf p}),
\ee
where
\be 
d_0({\bf p}) = {C \over 4\rho_s k^2 - 2 i M_0 \omega_m} = {CL^2\over
4\pi^2 \rho_s} {\lambda \over \lambda l^2 - 2\pi i m}, 
\ee
in which $\lambda = 2\pi^2 L_0^2/L^2$ as before. The subscript $0$ on the 
functional average indicates that it is taken using the action $S_0$ 
(which does not contain $\bf A$).
The replica index structure of the interaction vertex~(\ref{interaction})
is such that this diagonal property holds in all orders of perturbation
theory, so that the full propagator is also diagonal when ${\bf 
A}=0$, $\la\!\la \bd_{pq} \bph_{rs} \ra\!\ra_{{\bf A}=0} \propto \de_{pr} 
\de_{qs}$ . Then it immediately follows that the
contributions to $\mean{g}$ quadratic in $b$ disappear in the replica
limit: $\la\!\la (\bdg b)_{11} \ra\!\ra_{{\bf A}=0} = 
\la\!\la \bd_{i1} \bph_{i1} \ra\!\ra_{{\bf A}=0} \propto 
\de_{ii} = n \to 0$, and, similarly, $\la\!\la (b\bdg)_{11} \ra\!\ra_{{\bf 
A}=0} \to 0$.

The same argument applies to the first two terms in the 
expressions~(\ref{diagcurrents}) for the diagonal currents. Moreover, it 
is easy to see that the quartic terms there after averaging also contain 
at least one factor of $n$, in all orders of perturbation theory. Then it 
follows that 
\be
\limit \left\la\!\!\left\la I^{(2)}_{11}(x)\right\ra\!\!\right\ra = \limit 
\left\la\!\!\left\la I^{(1)}_{11}(x)\right\ra\!\!\right\ra = 0,
\ee
which proves the divergencelessness of the conductance, Eq.~(\ref{wicond}).

The bare propagator in real space
\be
d_0({\bf r};{\bf r'}) = {L\over 2\pi^2 C\rho_s} \sum_{l,m}
{\lambda \sin k_l x\sin k_l x'\over\lambda l^2 - 2\pi i m} 
e^{i \omega_m (\tau - \tau')}
\ee
has the dimensionful prefactor $L/\xi$.  The interaction vertex
$S_{\rm int}$~(\ref{interaction}) has the dimension $C\rho_s/L \propto
\xi/L$ by inspection.  Then in the $M$-th order of perturbation theory the
contribution to a general correlator of $2N$ bosonic operators in real
space, having $2M + N$ bare propagators in it, will be proportional to
$(L/\xi)^{M+N}$.  Therefore to order $L/\xi$, which is all we need to 
obtain the function $\Phi$ of Eq.~(\ref{meang}), we have to calculate 
only two correlators in the zeroth order of perturbation theory. 

The first correlator is 
\bea
\left\la\!\left\la(\bdg b)_{11}(b\bdg)_{11} 
\right\ra\!\right\ra_0 &=& \left\la\!\!\left\la\bd_{i1} \bph_{1j} 
\right\ra\!\!\right\ra_0 \left\la\!\!\left\la \bph_{i1} \bd_{1j} 
\right\ra\!\!\right\ra_0 \no\\
&=& d_0^2(x,\tau^+;x,\tau)
\eea
(we used Wick's theorem and dropped the
term disappearing in the replica limit), and the corresponding
contribution to the function $\Phi$ is 
\bea
&&{4\over\pi^4} \left( \left( \sum_{l,m} {\lambda e^{i\omega_m\eta}\over
\lambda l^2 - 2\pi i m} \right)^2 \right.\no \\ 
&&\left.\vphantom{\left( \sum_{l,m} {\lambda e^{i\omega_m\eta}\over
\lambda l^2 - 2\pi i m} \right)^2} + {1\over 2} \sum_{l,m_1,m_2} {\lambda 
e^{i\omega_{m_1}\eta} 
\over \lambda l^2 - 2\pi i m_1} {\lambda e^{i\omega_{m_2}\eta} \over
\lambda l^2 - 2\pi i m_2} \right), 
\eea
where $\eta$ is a positive infinitesimal. When the frequency summations
over $m$ are done, we obtain the first two terms in Eq.~(\ref{Phi}).  The
second correlator we need is 
\bea
&&\left\la\!\left\la (b\dx\bdg b)_{11}({\bf r}) 
(\bdg \dxprime b \bdg)_{11}({\bf r'}) \right\ra\!\right\ra_0 \br
= \dx\dxprime d_0({\bf r};{\bf r'}) d_0^2({\bf r'};{\bf r}), 
\eea
and the corresponding
contribution to $\Phi$ gives the last term in Eq.~(\ref{Phi}) (using
$f(x) - f(y) = f(x) f(-y)/f(x-y)$). 

The calculation of the variance of the conductance proceeds along the same
lines, except that the power counting and the replica limit leave us only
with two correlators to compute. The first one is 
\bea
%\lefteqn{
&&\left\la\!\left\la(\bdg b)_{11}({\bf r}) (b\bdg)_{22}({\bf r'}) + 
(b\bdg)_{11}({\bf r}) (\bdg b)_{22}({\bf r'}) \right\ra\!\right\ra_0 \no\\
&& = 2d_0({\bf r};{\bf r'})  d_0({\bf r'};{\bf r}),
\eea 
and the second is 
\be
\left\la\!\!\left\la (\bph_{21} \bph_{12})({\bf r}) 
(\bd_{12}\bd_{21})({\bf r'}) \right\ra\!\!\right\ra_0 = 
d_0^2({\bf r'};{\bf r}). 
\ee
The function $\widetilde{\Phi}$ is given then by
\bea
\widetilde{\Phi} \left(L/L_0\right) & = & {32 \rho_s^2 \over L^4} \intr \!
\intx'\!\intt' \left( 2 d_0^2({\bf r'};{\bf r})\right. \no \\
&& \left.+ d_0({\bf r};{\bf r'})  d_0({\bf r'};{\bf r}) \right).
\label{Phitilde3}
\eea
An expression of this form is quite standard in the universal conductance
fluctuation theories, see for example, Eq.~(5.3) in Ref.~\cite{xrs}. 
The only
difference is the anisotropic form of the diffusion propagator $d_0({\bf
r};{\bf r'})$. In momentum space we obtain
\be
\widetilde{\Phi} \left(L/L_0\right) = {2 \lambda^2\over \pi^4} \sum_{l,m}
\left( {2 \over |\lambda l^2 - 2\pi i m|^2} + {1 \over (\lambda 
l^2 - 2\pi i m)^2} \right). \label{Phitilde4}
\ee
Performing frequency summations we arrive at the
expression~(\ref{Phitilde}) [using $f(x) + f(-x) = -1$ and $f'(x) = f(x) +
f^2(x) = - f(x)f(-x)$]. 

The limiting values~(\ref{Phi1}--\ref{Phitilde2}) are obtained using the
asymptotic forms of the Bose function:
\be
f(x) \sim \left\{ \begin{array}{ll}
1/x, & x \to 0, \\
e^{-x}, & x \to +\infty, \\
-1, & x \to -\infty.
\end{array} \right.
\ee
In the 1D limit $L \gg L_0$ ($\lambda \ll 1$) the expression~(\ref{Phi})
reduces to 
\be
{4\over\pi^4} \left( {1\over 2} \sum_{l=1}^{\infty} {1\over l^4} - 
2 \sum_{l_1,l_2 = 1}^{\infty} {1\over l_1^2(l_1 + l_2)^2} \right).
\ee
These sums are easily evaluated, and we obtain Eq.~(\ref{Phi1}).
Similarly, the expression~(\ref{Phitilde}) in the same limit $L \gg L_0$
simplifies to $6\pi^{-4}\sum_l l^{-4} = 1/15$. In the opposite 2D chiral
metal limit $L \ll L_0$ ($\lambda \gg 1$) all the terms in Eq.~(\ref{Phi})
are exponentially small, and the largest of them gives Eq.~(\ref{Phi2}).
In contrast, the first term in Eq.~(\ref{Phitilde}) diverges in this limit
giving Eq.~(\ref{Phitilde2}).

The two terms in eq.~(\ref{Phitilde3}) correspond to two distinct 
correlation functions as above, or to two Feynman diagrams. 
Mathur~\cite{m} evaluated only the first of these, in the two limits 
$\lambda\rightarrow\infty$ and $\lambda\rightarrow0$. It happens to give 
the full contribution asymptotically as $\lambda\rightarrow\infty$. The 
remaining diagram is nonzero in general, and for $\lambda\ll 1$ 
contributes exactly $1/2$ as much as the first one, to give the total in 
eq.\ (\ref{Phitilde1}). 

%%%%%%%%%%%%%%%%%%%%%%%%%%%%%%%%%%%%
\section{Conclusion}
\label{conclusion}

In conclusion, we considered the directed network model of edge states. In
previous studies~\cite{bfz,m,grs} it was shown that in a continuum limit
this model can be mapped to a 1D quantum ferromagnetic spin chain.  Three
regimes, 2D chiral, 1D metallic, and 1D localized, separated by
smooth crossovers, have been identified for the model, and the universal
crossover functions for the mean and variance of the conductance have been
obtained for the crossover between 1D regimes in our previous
paper~\cite{grs}. In this paper we use spin-wave perturbation theory to
obtain the corresponding universal functions for the other crossover, 
between the 2D chiral and 1D metallic regimes. The results and their
asymptotic forms are given in Sec.~\ref{outline}, see
Eqs.~(\ref{Phi}--\ref{Phitilde2}), and were discussed there.

\section*{Acknowledgments}

This work was supported by NSF grants, Nos.\ DMR--91--57484 and
DMR--96--23181. The research of IG and NR was also supported in part by 
NSF grant No.\ PHY94-07194.

\end{document}